\documentclass[useAMS,usenatbib]{mn2e}

\usepackage{graphicx}
\usepackage{amsmath}
\usepackage{subfigure}
\usepackage{rotating}
\usepackage{url}

\title[Optimal SKA Dish Configuration using Genetic Algorithms]
  {Optimal SKA Dish Configuration using Genetic Algorithms}
\author[A. ~Gauci et al.]
  {~Adam ~Gauci$^1$, ~Kristian ~Zarb ~Adami$^2$, ~John ~Abela$^3$, ~Babak ~E. ~Cohanim$^4$\\
 $^1$Department of Intelligent Computer Systems, Faculty of ICT, University of Malta, Malta.\\
 $^2$Department of Physics, Faculty of Science, University of Malta, Malta.\\
 $^3$Department of Computer Information Systems, Faculty of ICT, University of Malta, Malta.\\
 $^4$Mission Design Group Leader, Draper Laboratory, 555 Technology Square, Cambridge MA 02139.\\
}

\date{Released 2011 Xxxxx XX}

\pagerange{\pageref{firstpage}--\pageref{lastpage}} \pubyear{2011}

\def\LaTeX{L\kern-.36em\raise.3ex\hbox{a}\kern-.15em
    T\kern-.1667em\lower.7ex\hbox{E}\kern-.125emX}

\begin{document}

\label{firstpage}

\maketitle

\begin{abstract}

The Square Kilometre Array (SKA) is a radio telescope designed to
operate between 70MHz and 10GHz. Due to this large bandwidth, the
SKA will be built out of different collectors, namely antennas and
dishes to cover the frequency range adequately. In order to deal
with this bandwidth, innovative feeds and detectors must be
designed and introduced in the initial phases of development.
Moreover, the required level of resolution may only be achieved
through a groundbreaking configuration of dishes and antennas. Due
to the large collecting area and the specifications required for
the SKA to deliver the promised science, the configuration of the
dishes and the antennas within stations is an important question.
This research builds on the work done before by \cite{cohanim04a},
\cite{hassan05} and \cite{grigorescu09} to further investigate the
applicability of machine learning techniques to determine the
optimum configurations for the collecting elements within the SKA.
This work primarily uses genetic algorithms to search a large
space of optimum layouts. Every genetic step provides a population
with candidate individuals each of which encodes a possible
solution. These are randomly generated or created through the
combination of previous encodings. In this study, a number of
fitness functions that rank individuals within a population of
dish configurations are investigated. The UV density, connecting
wire length and power spectra are considered to determine a good
dish layout.

\end{abstract}

\begin{keywords}
SKA, radio telescope, machine learning, evolutionary programming,
genetic algorithms
\end{keywords}

\section{Introduction} \label{IntroductionSection}

The SKA will be an instrument through which major scientific
discoveries are to be made. Although the construction will follow
a phased approach, phase 1 of the SKA is already a formidable
instrument and will undoubtedly shed light on the evolutionary
stages of the universe from the epoch of reionisation as well as
improve our understanding of gravity through the detection of
binary and millisecond pulsars \citep{garrett10}.

Dishes and antenna arrays will, using state of the art receivers,
provide unprecedented sensitivity between 70MHz and 10GHz
\citep{garrett10}. The required resolving power unavoidably
dictates an enormous spatial extent ($\approx$ 3000km) in the
initial phase and will cost around 500M Euro \citep{dewdney10}. A
pioneering design  minimising infrastructure, networking and other
costs whilst still achieving the desired specifications is of
importance both in the construction phase, but more importantly in
the maintenance and running costs of the telescope.
\cite{grigorescu09} estimate that 100M Euro will most likely be
allocated to cabling and trenching that connects the stations
together.

In this study, the applicability of Genetic Algorithms (GA) to
determine the most optimum configurations for the dish array is
investigated. Such evolutionary programming approaches are based
on Darwin's theory of natural selection in which the fittest
individuals from each population survive and generate offspring
chromosomes that encode configurations which are closer to the
optimum solution. In Section \ref{geneticAlgorithmsSection}, an
introduction to GAs is presented while in Section
\ref{dishConfigurationSection}, the work done for dish array
optimisation is discussed. Following details on the implemented
genetic operators and fitness functions, various cases together
with the obtained results are presented in Section
\ref{dishResultsSection} and Section
\ref{dishConfigurationPhase2ResultsSection}. Some conclusions are
drawn in Section \ref{conclusionSection}.

\section{Genetic Algorithms}\label{geneticAlgorithmsSection}
Genetic Algorithms (GAs) are search heuristics that follow the
natural process of evolution to determine the most fit hypothesis
from a pool of possible solutions. Unlike other search techniques
that adopt a brute-force or iterative strategy, GAs combine parts
of the best know solutions to try and create better encodings.
This evolutionary programming methodology that uses both mating
and mutation to create better chromosomes was pioneered by John
Holland around the mid 1960's \citep{holland05}. Since then, GAs
have been used for a wide range of applications where an optimized
solution with a large number of parameters is required.

Chromosomes that represent valid hypothesis need to be encoded as
streams of data that can be processed by the algorithm. A pool of
such encodings is referred to as the population and the GA
progresses by updating this set of solutions. In each generation,
new individuals are created randomly or through genetic operators
such as crossover and mutation that recombine or mutate parent
chromosomes respectively. Parent hypothesis from which offsprings
are created, are selected according to a probability function.

In each generation step, all solutions are ranked by a fitness
function and the population is updated to include the best
individuals. The process is repeated until the algorithm stalls
and no improvement in the fitness is detected with further
processing. In this work, each chromosome represented a
configuration and stored the dish locations.

As discussed by \cite{mitchell97}, GAs search through a large
space to find the solution that maximises the fitness function.
With the adopted approach, the algorithm is less likely to
converge towards a local minimum since the operators can replace
parent encodings with completely different offsprings. As the
algorithm progresses, one must make sure that a group of good and
similar encodings will not replicate and dominate the population.

\section{Configuration} \label{dishConfigurationSection}

\begin{figure}
\centering
\includegraphics[width=80mm]{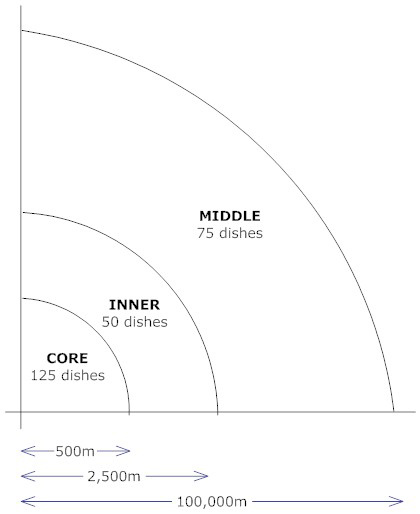}
\caption {SKA phase 1 dish layout.} \label{dishPhase1LayoutFigure}
\end{figure}

The specification document for phase 1 SKA specifies that 250
parabolic dishes each 15m in diameter will be installed over a
100km radius region \citep{dewdney10}. 125 core stations (50\%)
will be fixed in the central 500m radius. The inner region and
middle region will extend over a radius of 2,500m and 100,000m and
will contain 50 (20\%) and 75 (30\%) antennas respectively. Figure
\ref{dishPhase1LayoutFigure} shows this layout.

In this study, a search for the optimum configuration that
maximises the uniformity of the UV density distribution while
keeping the connecting wire length to a minimum was conducted. The
goal is to position the dishes in such a way as to obtain a flat
uv distribution with points spread uniformly across the uv-plane
\citep{cohanim04b}. A regular gridded mask representing the domain
over which the dishes can be positioned, was initially generated.
The initial encodings with possible configurations were then
created with antenna locations chosen randomly from such a grid.
The algorithm was let to evolve for a number of generations until
no improvement in the fitness was detected and all encodings in
the final population were similar. Due to the required number of
dishes and receiver distribution, the initial population could not
be biased with previously known good encodings.

\subsection{Chromosome structure and genetic operators}\label{dishChromosomeStructureandGeneticOperators}

\begin{figure}
\centering
\includegraphics[width=80mm]{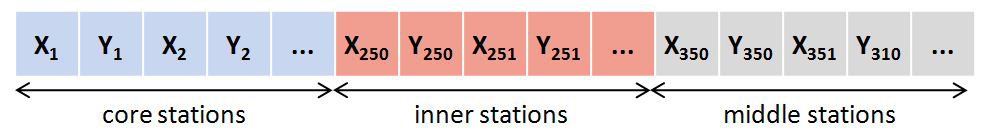}
\caption {Dish configuration chromosome structure.}
\label{dishChromosomeFigure}
\end{figure}

Encodings that represented different configurations were created
each one storing the $x$ and $y$ coordinates of the dish
locations. As shown in Figure \ref{dishChromosomeFigure}, each
chromosome stored the mapping of dishes on the domain grid as a
series of 500 integers. An identification number was also
associated and stored with each encoding. This allowed the
properties and status of each chromosome to be monitored and
saved.

\begin{figure}
\centering
\includegraphics[width=80mm]{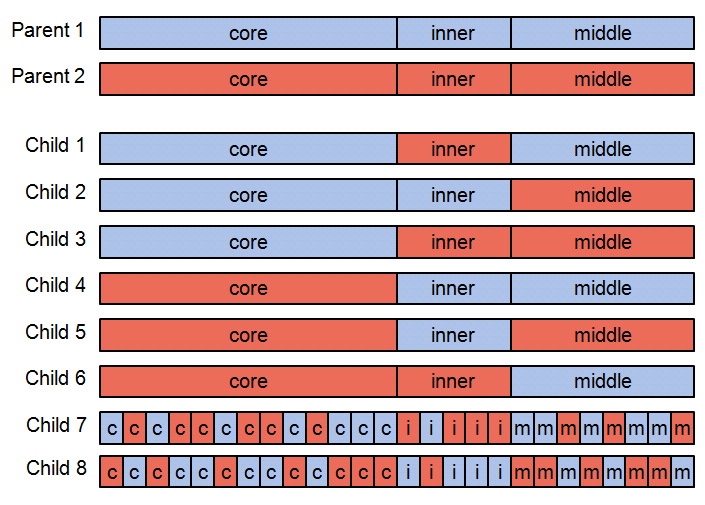}
\caption {Offsprings created by the dish configuration crossover
operator.} \label{dishCrossoverFigure}
\end{figure}

The crossover operation was designed to produce offsprings whose
genes encode combinations of the core, inner and middle region.
From every pair of parent chromosomes, after all gene combinations
have been carried out, six new individuals were created. If the
core, inner and middle regions of the first parent were
represented by \texttt{C1 I1 M1}, and the second parent was made
from \texttt{C2 I2 M2}, encodings with \texttt{C1 I2 M1},
\texttt{C1 I1 M2}, \texttt{C1 I2 M2}, \texttt{C2 I1 M1},
\texttt{C2 I1 M2} and \texttt{C2 I2 M1} were generated. To create
more offspring by combining existing chromosomes, two more
encodings were generated according to a randomly generated binary
vector. In particular, dish positions corresponding to one were
taken from the first parent while positions corresponding to zero
were taken from the second parent. The binary vector was then
bitwise inverted and the same procedure was repeated to obtain one
more configuration. Since parent chromosomes may have common dish
locations, the last two generated offsprings  where checked by a
gene repair function to ensure that all 250 locations were
distinct. This crossover process is graphically shown in Figure
\ref{dishCrossoverFigure}.

The mutation operator was implemented such as to alter the
positions of randomly selected dish locations. This allowed the
algorithm to keep searching and to consider closely related
encodings in the multidimensional search space. Although most
parts of the chromosomes remained unchanged after mutation, the
shifting of some of the dishes to a new location prevented the
algorithm from converging onto a local maximum. In particular, a
vector with 250 random values between 0 and 1 was populated.
Indices of locations to which a number less than 0.2 was assigned
were identified and new dish locations to the corresponding
positions were determined. Chromosomes created by mutation were
processed by the gene repair function to ensure that no location
had more than one dish assigned to it.

\subsection{Fitness functions}\label{dishConfigurationFitnessFunctionsSubSection}

\subsubsection{UV density distribution
fitness}\label{dishUvDensityDistributionSubSubSection}

In order to ensure that the genetic algorithm converged towards a
solution that maximised UV coverage, the density map was computed
from every unique pair of dishes by equation \ref{uvEq}.

\begin{equation}\label{uvEq}
\left(\begin{array}{c}
u_{i,j}\\
v_{i,j}\\
w_{i,j}
\end{array}\right)=\frac{1}{\lambda}\left(\begin{array}{ccc}
x_{i} - x_{j}\\
y_{i} - y_{j}\\
z_{i} - z_{j}
\end{array}\right)
\end{equation}

Due to the nature of the dish array, $N(N-1)/2$ number of unique
points (where $N$ is the number of dishes) were generated. In
certain test cases, the full coverage of the telescope after
taking into consideration the rotation of the earth was also
computed. As discussed by \cite{segransan07}, such projection can
be determined by equation \ref{uvHourEq}. Here, $h$ represents the
hour angle and $\delta$ is the source declination.

\begin{table*}
\begin{equation}\label{uvHourEq}
\left(\begin{array}{c}
u_{i,j}\\
v_{i,j}\\
w_{i,j}
\end{array}\right)=\frac{1}{\lambda}\left(\begin{array}{ccc}
\sin(h) & \cos(h) & 0\\
-\sin(\delta)\cos(h) & \sin(\delta)\sin(h) & \cos(\delta)\\
\cos(\delta)\cos(h) & -\cos(\delta)\sin(h) & \sin(\delta)
\end{array}\right)\left(\begin{array}{c}
x_{i}-x_{j}\\
y_{i}-y_{j}\\
z_{i}-z_{j}
\end{array}\right)
\end{equation}
\end{table*}

In order to be able to compare the resulting outputs, in this work
the declination was always set to $90^\circ$ to represent a radio
object at the celestial north pole while the hour angle was set to
range from $0^\circ$ to $345^\circ$ at $15^\circ$ intervals.

Since the computation of the distance between all baselines
becomes prohibitively expensive very quickly, we followed the work
published by \cite{cohanim04b} and \cite{cohanim04a}. In
particular, the nominal grid point closest to each UV point was
determined and flagged. An analysis of the non-matched nominal
points gave an indication of the distribution of the configuration
and hence a measure of fitness. Ideally, the majority of nominal
grid points would be flagged by at least one UV point.

\begin{figure}
\centering
\includegraphics[width=80mm]{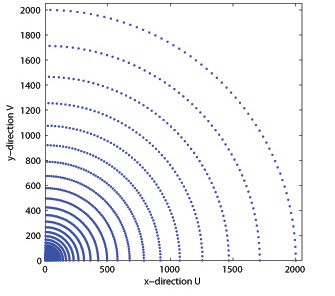}
\caption {UV nominal grid quadrant.} \label{nominalGridFigure}
\end{figure}

Due to its size and current vision, the SKA will be a log based
structure. As shown in Figure \ref{nominalGridFigure}, a log
distribution for the nominal grid was decided to be used. The goal
of the GA was then set to minimise the fitness function, i.e. the
percentage of non-matched nominal grid points. As in
\cite{cohanim04a}, these were calculated using equation
\ref{fUvEq}.

\begin{equation}\label{fUvEq}
f_{UV} = \frac {N_{total} - N_{matched}} {N_{total}}
\end{equation}

Here, $N_{total}$ is the total number of points in the nominal
grid and $N_{matched}$ is the total number of matched points. The
numerator equates to the percentage of grid points that were not
matched with any UV point.

Fitness evaluation of every individual required an efficient
calculation of the UV density distribution as well as the mapping
onto the nominal grid for a large number of chromosomes. We
selected to use a k-dimensional tree representation of the nominal
grid which needed to be computed only once and could be then
stored in memory. The nearest point to every position
encoded could then be determined by traversing the constant binary
tree data structure. Since the nominal grid was defined over two
dimensions, each non-leaf node represented a perpendicular
hyperplane that divided the space into two subspaces. The left
subtree pointed to other nodes on the left while the right subtree
represented points to the right.

\subsubsection{Logarithmic wire length
fitness}\label{dishWireLengthLogFitnessSubSubSection}

Various approaches that attempt to compute an accurate cost and
minimise the required length of cable to connect the dishes
together, have been presented. \cite{grigorescu09} provide a set
of algorithms that also take into account trenching, as well as
connection costs to optimise a telescope layout infrastructure. In
\cite{cohanim04a}, the single linkage algorithm is used. Here, to
determine the shortest sequence that connects all vertices
together, the Kruskal Minimum Spanning Tree (MST) algorithm
\citep{cormen01} was used.

Throughout this work, a cable with unit cost per unit length that
connects all dishes in the core, inner and middle regions, was
assumed. Dish locations were connected in such a way as to create
an undirected graph in which edges (connections) between each
vertex (dishes) had no particular direction. The weight of every
edge was taken to correspond to the Euclidian distance between the
two connecting nodes. The MST algorithm was then use.

Since the UV density fitness corresponds to a percentage ranging
from 0 (optimum UV distribution) to 1 (worst case), a normalizing
function that allows the computed wire length to be compared and
added with the resulting UV fitness, was required. A log based
approach was initially adopted and the cable length fitness was
computed by equation \ref{fWireLogEq}.

\begin{equation}\label{fWireLogEq}
f_{WireLog} = 1 - log_{10} \left ( \frac{1}{wire length} \right )
\end{equation}

\begin{figure}
\centering
\includegraphics[width=80mm]{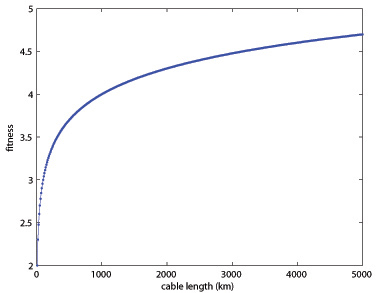}
\caption {Log scale cable length fitness ($f_{WireLog}$).}
\label{logCableLengthFigure}
\end{figure}

The wire length is given in kilometers. Figure
\ref{logCableLengthFigure} shows the fitness values for cable
lengths between 0 and 5000km.

\subsubsection{Stepwise wire length
fitness}\label{dishWireLengthStepFitnessSubSubSection}

Since the majority of chromosomes were found to have a cable
length of about 1000km, a stepwise function that linearly varies
the output between 0.1 and 0.8 for wire lengths between 900km and
1300km was implemented. More specifically, the wire length fitness
in this case was computed by equation \ref{fWireStepEq}.

\begin{equation}\label{fWireStepEq}
f_{WireStep} =
\begin {cases}
0 &\text{if $0 \leq wirelength < 100$;}\\
0.05 &\text{if $100 \leq wirelength < 500$;}\\
0.1 &\text{if $500 \leq wirelength < 900$;}\\
0.1 \rightarrow 0.8 &\text{if $900 \leq wirelength < 1300$;}\\
0.8 &\text{if $1300 \leq wirelength < 1400$;}\\
0.9 &\text{if $1400 \leq wirelength < 1500$;}\\
1 &\text{otherwise;}\\
\end{cases}
\end{equation}

\begin{figure}
\centering
\includegraphics[width=80mm]{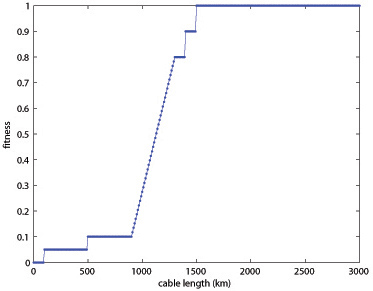}
\caption {Stepwise cable length fitness ($f_{WireStep}$).}
\label{stepCableLengthFigure}
\end{figure}

In this way, the algorithm could accurately assign and rank
individuals. Figure \ref{stepCableLengthFigure} depicts this
stepwise variation with wire length more clearly.

\subsubsection{Wire length penalty fitness}\label{dishWireLengthPenaltyFitnessSubSubSection}

\begin{figure}
\centering
\includegraphics[width=80mm]{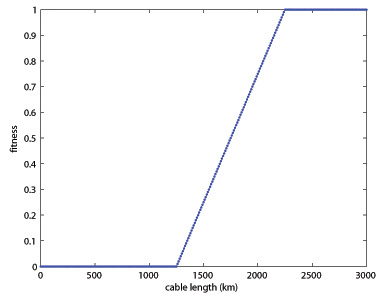}
\caption {Penalty cable length fitness ($f_{WirePenalty}$).}
\label{penaltyCableLengthFigure}
\end{figure}

Further tests suggested that a wire length penalty approach may be
more effective. Individuals encoding dish locations that could be
connected by a cable length of less than 1250km, were not
penalised. Chromosomes with a minimum wire length greater than
2250 were highly discouraged through a fitness assignment of 1.
Intermediate cable lengths were given a weighting which varied
linearly as described by equation \ref{fWirePenaltyEq}. This
variation of wire length fitness is presented in Figure
\ref{penaltyCableLengthFigure}. The threshold values used were
determined after noting the results obtained in previous runs. The
main advantage of this approach was that it directed the search
towards solutions with a good UV coverage and penalized encodings
that have a wire length above the norm. All encodings with a cable
length of less than 1250km were treated equally and the fitness
was taken to depend solely on the UV distribution.

\begin{equation}\label{fWirePenaltyEq}
f_{WirePenalty} =
\begin {cases}
0 &\text{if $0 \leq wirelength < 1250$;}\\
0 \rightarrow 1 &\text{if $1250 \leq wirelength < 2250$;}\\
1 &\text{otherwise;}\\
\end{cases}
\end{equation}

As discussed in subsequent sections, in order to compare the
results obtained in this study with a generic configuration,
dishes in the middle region were clustered together. This group
formation naturally minimised the wire length and to account for
these  encodings, another wire penalty fitness function with lower
thresholds was defined. This is formally defined by equation
\ref{fWirePenaltyLowEq} below.

\begin{equation}\label{fWirePenaltyLowEq}
f_{WirePenaltyLow} =
\begin {cases}
0 &\text{if $0 \leq wirelength < 300$;}\\
0 \rightarrow 1 &\text{if $300 \leq wirelength < 450$;}\\
1 &\text{otherwise;}\\
\end{cases}
\end{equation}

\subsubsection{Power spectrum fitness}\label{dishPowerSpectrumFitnessSubSubSection}

\begin{figure}
\centering
\includegraphics[width=80mm]{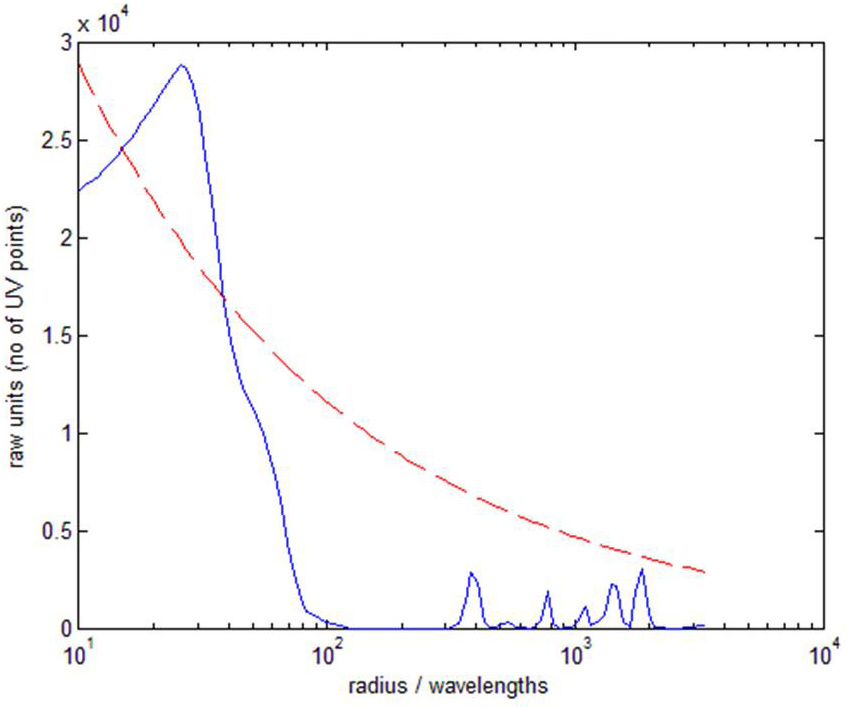}
\caption {Raw angular power spectrum (blue) and a log decaying
curve used as reference for fitness calculation (red).}
\label{fitnessPowerSpectrumFigure}
\end{figure}

Any improvement gained through the introduction of power spectra
calculation as part of the fitness function, was also
investigated. Studies such as \cite{parsons11} provide detailed
algorithms of how to compute power spectrum. However, for this
study, work done by \cite{green07} was followed to determine the
raw angular power spectrum from the UV-plane. In particular, the
number of UV points that coincided with log spaced annuli of width
equal to the restricted zone diameter of the dishes, was
determined. The resulting data series was divided by a log
decaying curve and a mean value was computed to obtain a measure
of fitness proportional to the distance between the two curves
($f_{PowerSpectrum}$). A typical raw angular power spectrum and
the considered ideal curve are shown in Figure
\ref{fitnessPowerSpectrumFigure}.

As the GA progressed, the fitness of individuals in each
population were computed in parallel. The algorithm was left to
evolve until it stalled and there was very limited improvement
with subsequent processing. As discussed in Section
\ref{dishResultsSection} below, runs using various combinations of
the above mentioned fitness criteria were conducted.

\section{Results for SKA Phase 1}
\label{dishResultsSection}

An analysis of how the optimum configuration changes with
different fitness functions, population sizes, and criteria for
selecting individuals for subsequent generations, was carried out.
In the following subsections the results obtained for different
cases are presented.

\subsection{Case 1 - GA with UV and log scaled wire length fitness}
\label{dishConfigurationCase1ResultsSubSubSection}

As a first test run, the genetic algorithm was set with an initial
population of 1024 random chromosomes. For each individual, the
overall fitness was calculated by equation \ref{fDish1Eq}.

\begin{equation}\label{fDish1Eq}
f_{dish1} = f_{UV} + f_{WireLog}
\end{equation}

\begin{figure}
\centering
\includegraphics[width=80mm]{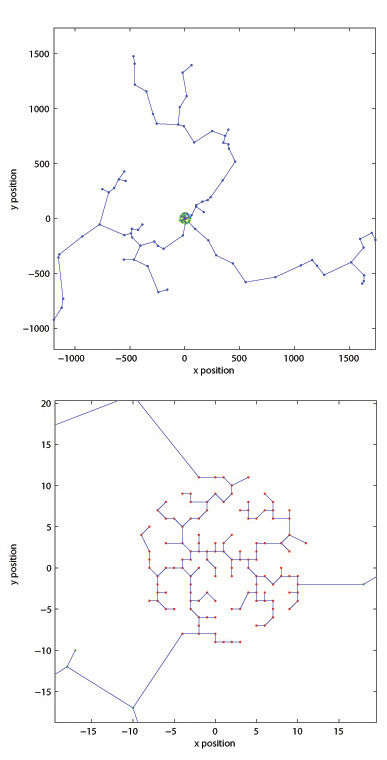}
\caption {Full (top) and zoomed (bottom) dish configuration with
shortest wire connecting the middle (blue), inner (green) and core
(red) regions for Case 1.} \label{dishCase1WireFigure}
\end{figure}

\begin{figure}
\centering
\includegraphics[width=80mm]{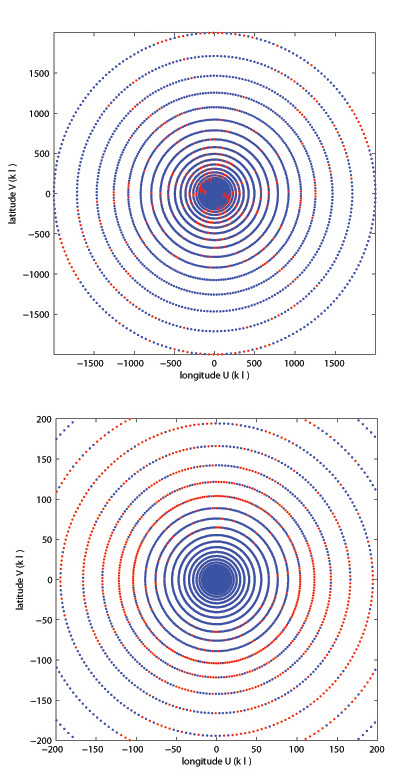}
\caption {Mapping of the UV density distribution onto the nominal
grid for the full array (top) and core region (bottom) showing the
matched (blue) and unmatched (red) points for Case 1.}
\label{dishCase1UvNominalFigure}
\end{figure}

Subsequent generations were created after selecting the fittest
1024 individuals from a pool of 4096 that consisted of 1024
chromosomes created by mutation, 2048 chromosomes created by
crossover and 1024 new randomly generated chromosomes. The initial
population had a mean and minimum fitness of 4.788 and 4.73
respectively. After 119 generations, the average fitness reduced
to 4.421 and the most optimum individual had a fitness of 4.414. A
plot of the resulting dish positions together with the computed
wire length is presented as Figure \ref{dishCase1WireFigure}. The
corresponding mapping of the UV density distribution onto the
nominal plane is shown in Figure \ref{dishCase1UvNominalFigure}.

\subsection{Case 2 - GA with weighted UV and stepwise cable length fitness}
\label{dishConfigurationCase2ResultsSubSubSection}

In the second case, the UV coverage and wire length were given a
weighting of 60\% and 40\% respectively as defined in equation
\ref{fDish2Eq}. Experimenting with different weighting schemes
allow the stakeholders to have a better understanding of the
tradeoffs between performance and cost.

\begin{equation}\label{fDish2Eq}
f_{dish2} = (0.6 \times f_{UV}) + (0.4 \times f_{WireStep})
\end{equation}

The initial population size was set to 1024. Individuals for
subsequent populations were selected from a pool of 1024
chromosomes generated through mutation, 2048 offsprings generated
by crossover and another 1024 random encodings. The highest
ranking chromosomes were also considered for migration into the
next population.

\begin{figure}
\centering
\includegraphics[width=80mm]{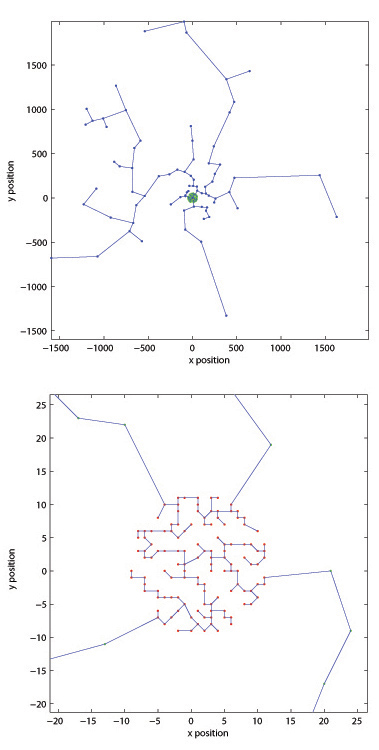}
\caption {Full (top) and zoomed (bottom) dish configuration with
shortest wire connecting the middle (blue), inner (green) and core
(red) regions for Case 2.} \label{dishCase2WireFigure}
\end{figure}

\begin{figure}
\centering
\includegraphics[width=80mm]{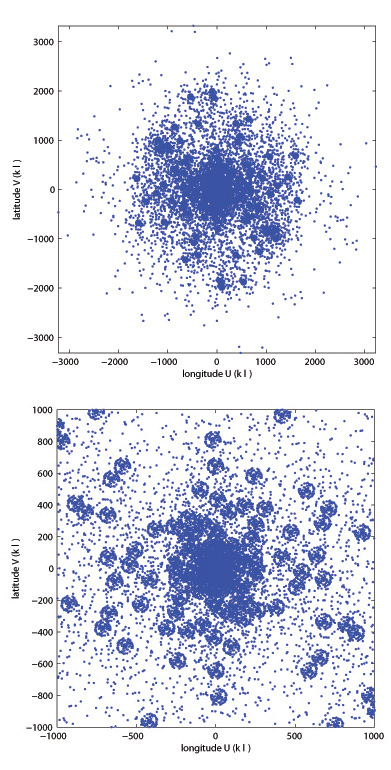}
\caption {UV density distribution for the full array (top) and
core region (bottom) for Case 2.} \label{dishCase2UvFigure}
\end{figure}

After the first few iterations, the percentage of randomly generated
chromosome rapidly decreased to zero. The selection of individuals
generated through mutation also decayed with time. The strongest
genes were created through crossover and the algorithm converged
after 111 iterations. Figure \ref{dishCase2WireFigure} shows the
final dish locations and wire length while the UV distribution is
presented in Figure \ref{dishCase2UvFigure}. This had a UV density
fitness of 0.67354 and wire length of 724.74km resulting in $f =
(0.6 \times 0.67354) + (0.4 \times 0.1) = 0.4441$.

\subsection{Case 3 - GA with UV and cable length penalty fitness}
\label{dishConfigurationCase3ResultsSubSubSection}

In this case, the input to the GA consisted of an initial
population with 4096 chromosomes which encoded random positions
for 250 dishes as defined in Section
\ref{dishChromosomeStructureandGeneticOperators}. At each step,
parent chromosomes were selected from the population to generate
4096 new offsprings through mutation and another 8192 new
individuals from crossover. The fitness of these new encodings as
well as another 4096 randomly generated individuals were combined
with the scores of the previous population and ranked to determine
the fittest 4096 entries. These were selected for the next cycle
and the process was restarted. In particular, the fitness was
computed by equation \ref{fDish3Eq}.

\begin{equation}\label{fDish3Eq}
f_{dish3} = f_{UV} + f_{WirePenalty}
\end{equation}

\begin{figure}
\centering
\includegraphics[width=80mm]{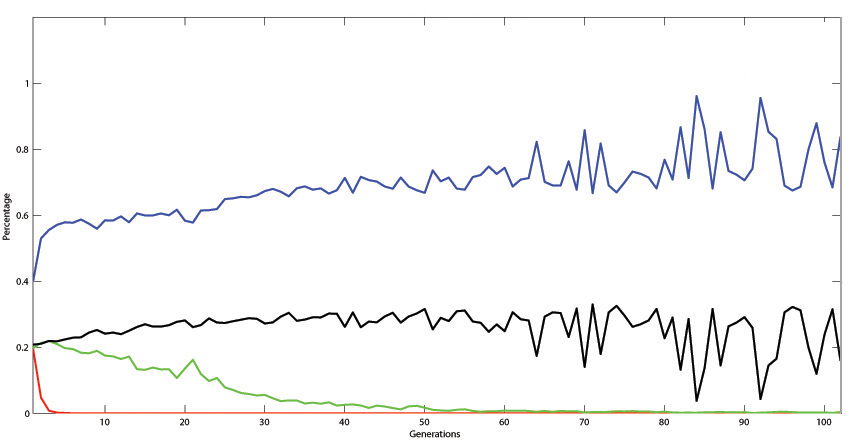}
\caption {Percentage of elite (black), crossover (blue), mutation
(green) and random (red) chromosomes selected for each population
for Case 3.} \label{dishCase3EvolutionFigure}
\end{figure}

\begin{figure}
\centering
\includegraphics[width=80mm]{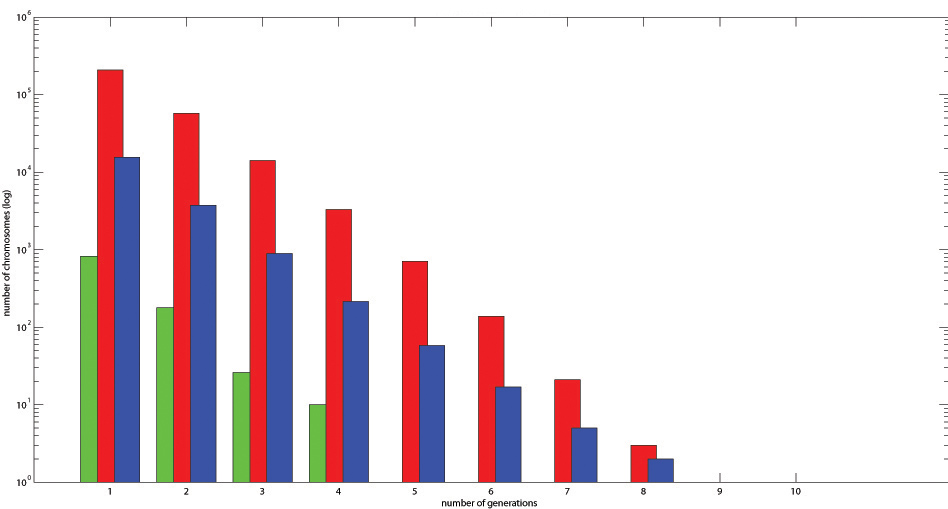}
\caption {Lifetime of crossover (red), mutation (blue) and random
(green) chromosomes for Case 3.}
\label{dishCase3ChromosomeLifetimeFigure}
\end{figure}

Figure \ref{dishCase3EvolutionFigure} gives an indication of the
percentage of elite, crossover, mutation and random chromosomes
selected at each generation. As expected, after the first few
iterations, the genetic operators produced individuals with
improved fitness and the algorithm progressed by continuously
choosing offsprings generated through crossover. Randomly
generated individuals became phased out and soon resulted to have
a lower fitness than the new offspring generated through the
combination of chromosomes already in the population. Figure
\ref{dishCase3ChromosomeLifetimeFigure} shows the typical lifetime
for crossover chromosomes, mutation chromosomes and randomly
created individuals before they got replaced by fitter members.
Encodings generated by the implemented genetic operators proved to
have a longer lifetime than random chromosomes. This speeded up
the convergence of the algorithm as well as permitted the
generation of fitter configurations.

\begin{figure}
\centering
\includegraphics[width=80mm]{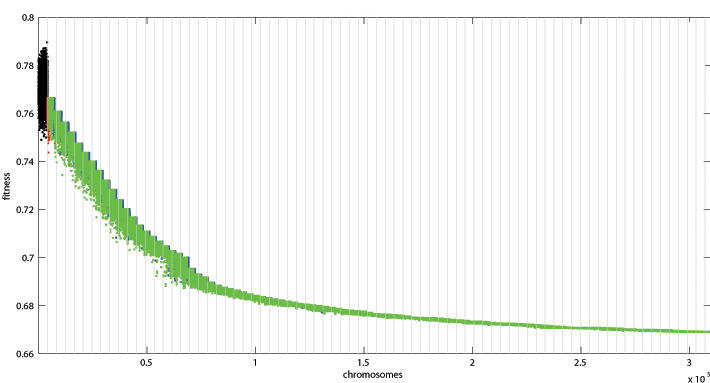}
\caption {Fitness for the initial individuals (black), random
chromosomes (red) and offsprings generated by the mutation (green)
and crossover (blue) operators for Case 3.}
\label{dishCase3FitnessFigure}
\end{figure}

\begin{figure}
\centering
\includegraphics[width=80mm]{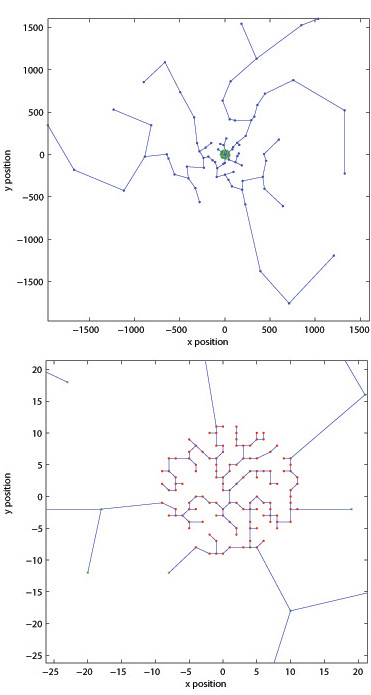}
\caption {Full (top) and zoomed (bottom) dish configuration with
shortest wire connecting the middle (blue), inner (green) and core
(red) regions for Case 3.} \label{dishCase3WireFigure}
\end{figure}

\begin{figure}
\centering
\includegraphics[width=80mm]{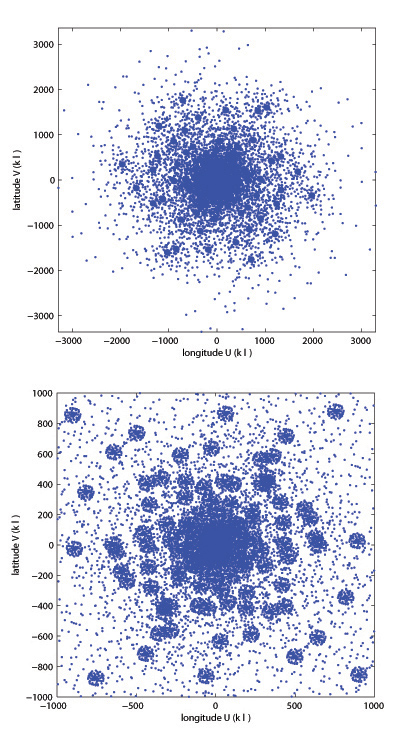}
\caption {UV density distribution for the full array (top) and
core region (bottom) for Case 3.} \label{dishCase3UVFigure}
\end{figure}

Figure \ref{dishCase3FitnessFigure} shows how the fitness improved
as the algorithm progressed. Figure \ref{dishCase3WireFigure}
presents a rendering of the fittest encoding after 102
generations. Dishes in the middle, inner and core regions are
shown in blue, green and red respectively. The UV density
distribution percentage was 0.66825 while the minimum wire length
computed by the MST algorithm was found to be 815.12km. Full and
zoomed versions of the UV distribution calculated from all dish
positions is presented in Figure \ref{dishCase3UVFigure}.

\subsection{Case 4 - GA considering randomly oriented grouped outer dishes with UV and cable length penalty fitness}
\label{dishConfigurationCase4ResultsSubSubSection}

In this case, the dish positioning and genetic functions were
modified so that configurations had the elements in the middle
region grouped in small random clusters of 3 to 8 dishes each.
Elements were positioned in a circular, triangular or linear
fashion and were given a random orientation. Since dishes were not
randomly scattered, the required cable length was expected to be
less and the fitness function defined by equation \ref{fDish4Eq}
was used.

\begin{equation}\label{fDish4Eq}
f_{dish4} = f_{UV} + f_{WirePenaltyLow}
\end{equation}

The crossover function used in the previous cases could still be
used since the middle region of all chromosomes had the exact same
number of elements. Genes from any two parents could be swapped
and still generate valid offsprings. However, the mutation
operator had to be redefined. If a dish within the middle region
was selected for mutation, a new position and shape for the entire
group had now to be determined. Since each encoding could have a
different number of groups with different number of dishes,
further logic had to be performed before randomising the
chromosome. Apart from the chromosome id, an integer with numerals
that corresponded with the number of dishes in each group was also
stored for each chromosome. Consecutive $(x, y)$ coordinates could
then be read until the require group of stations was found.

\begin{figure}
\centering
\includegraphics[width=80mm]{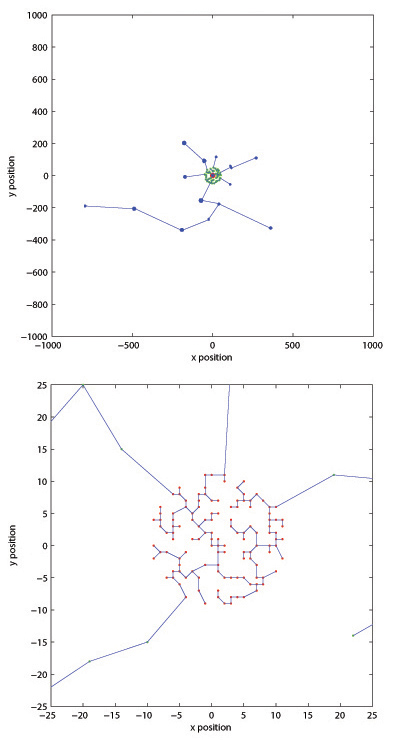}
\caption {Full (top) and zoomed (bottom) dish configuration with
shortest wire connecting the middle (blue), inner (green) and core
(red) regions for Case 4.} \label{dishCase4WireFigure}
\end{figure}

\begin{figure}
\centering
\includegraphics[width=80mm]{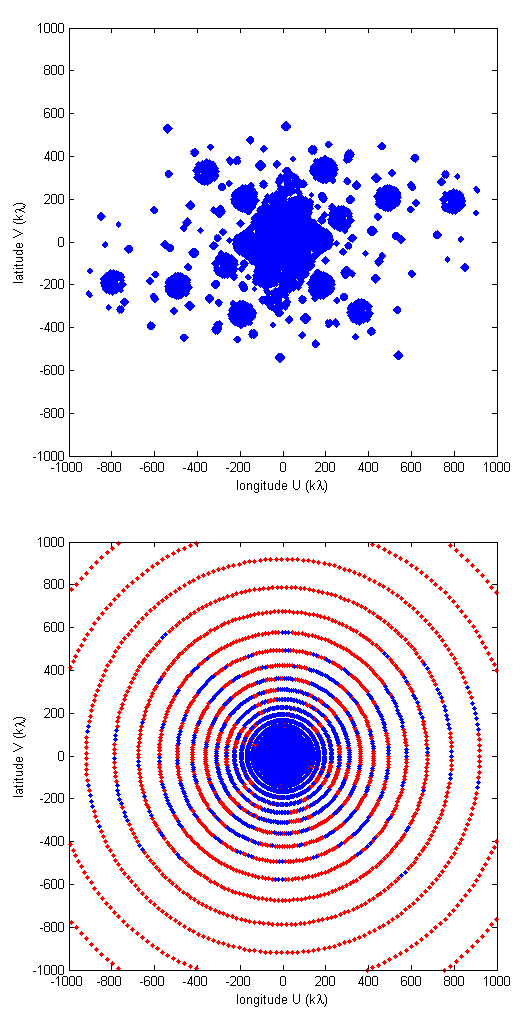}
\caption {UV density distribution (top) and mapping onto the
nominal grid (bottom) showing the matched (blue) and unmatched
(red) points for Case 4.} \label{dishCase4UVFigure}
\end{figure}

Figure \ref{dishCase4WireFigure} and Figure
\ref{dishCase4UVFigure} show the resulting configuration and the
connecting wire respectively. Although the UV density distribution
corresponds to a fitness of 0.77214 and a large number of nominal
grid points are unmatched, the clustering of dishes allowed for a
short cable length of 154.46km. For this run, the algorithm was
made to work on a population of 1024 chromosomes and evolved for
102 generations before it stalled.

\subsection{Case 5 - GA considering grouped outer dishes in a circular orientation with UV and cable length penalty fitness}
\label{dishConfigurationCase5ResultsSubSubSection}

\begin{figure}
\centering
\includegraphics[width=80mm]{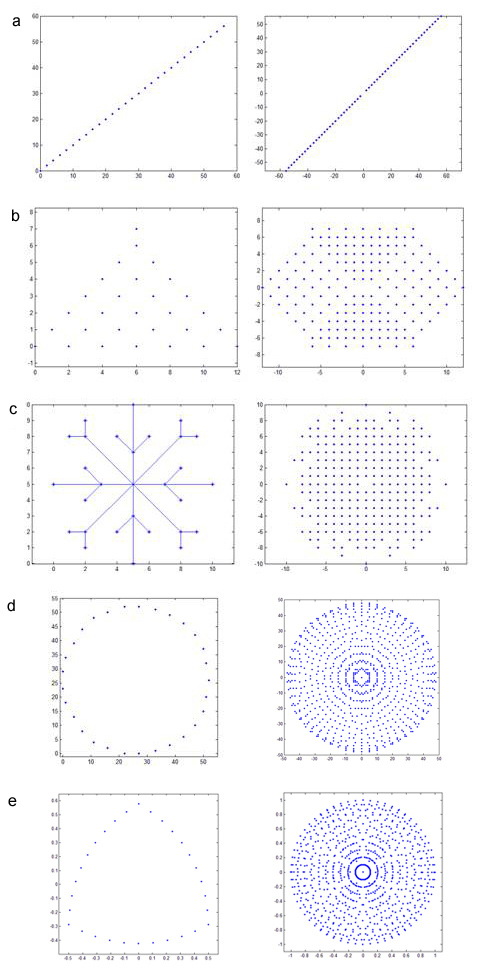}
\caption {Orientation of 29 dishes (right) and the corresponding
UV distribution (left) when placed in a straight line (a),
triangle (b), snowflake (c), circular (d), and reuleaux triangle
(e), configurations for Case 5.} \label{dishCase5ShapeUVFigure}
\end{figure}

Although dishes in the middle region were grouped as described for
the previous case, clustered elements were now only positioned and
oriented in a constant configuration. As shown in Figure
\ref{dishCase5ShapeUVFigure}, the corresponding UV distributions
for dishes positioned in a straight line, in a triangle, as a
snowflake, in a circular pattern and in a reuleaux triangle
orientation, were initially determined. In each case, 29 to 30
elements were used to render well the required shapes.For a small
number of elements, the configurations and the UV coverage of the
circular and reuleaux orientations are very similar and the
algorithm was set to distribute the dishes in the groups as such.

\begin{figure}
\centering
\includegraphics[width=80mm]{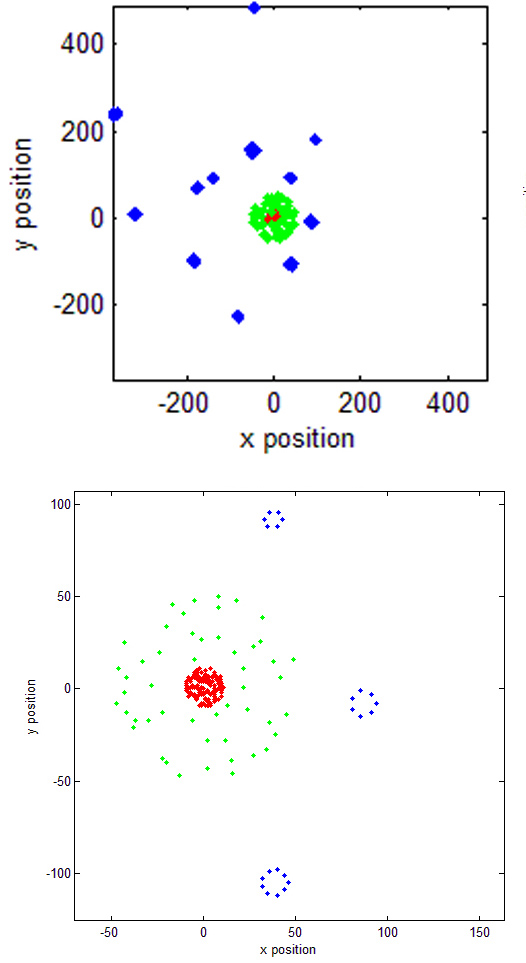}
\caption {Full (top) and zoomed (bottom) dish configuration
showing the middle (blue), inner (green) and core (red) regions
for Case 5.} \label{dishCase5PositionsFigure}
\end{figure}

\begin{figure}
\centering
\includegraphics[width=80mm]{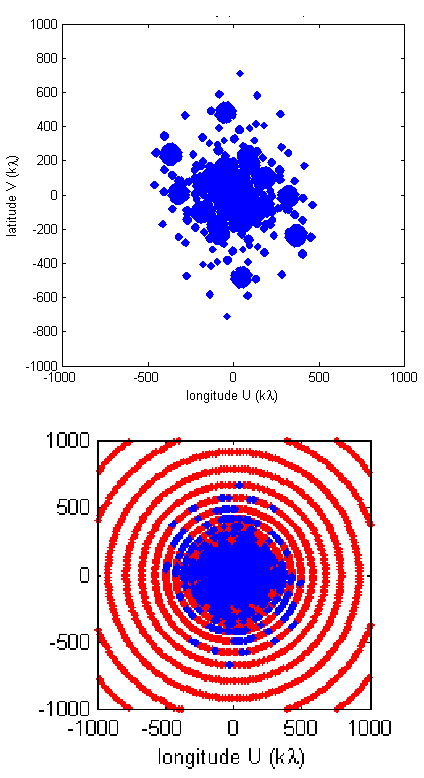}
\caption {Mapping of the 24 hour UV density distribution onto the
nominal grid for the full (top) and core region (bottom) showing
the matched (blue) and unmatched (red) points for Case 5.}
\label{dishCase5UVFigure}
\end{figure}

\begin{figure}
\centering
\includegraphics[width=60mm]{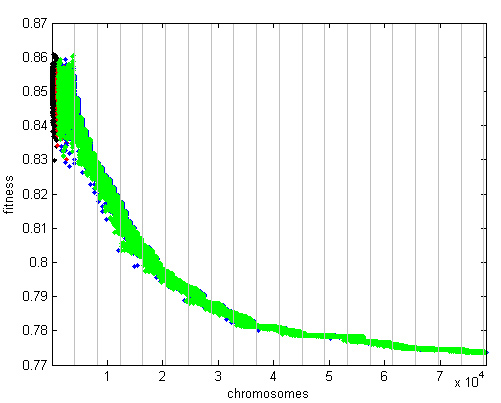}
\caption {Fitness for the initial individuals (black), random
chromosomes (red) and offsprings generated by the mutation (green)
and crossover (blue) operators for Case 5.}
\label{dishCase5FitnessFigure}
\end{figure}

Figure \ref{dishCase5PositionsFigure} shows one of the resulting
configurations after letting the GA run for 102 generations. The
corresponding UV pattern and UV mapping onto the nominal grid are
shown in Figure \ref{dishCase5UVFigure} while Figure
\ref{dishCase5FitnessFigure} shows the constant decay in fitness
with generations. Here the UV fitness and wire length resulted to
be 0.777321 and 120.94km respectively.

\begin{figure}
\centering
\includegraphics[width=80mm]{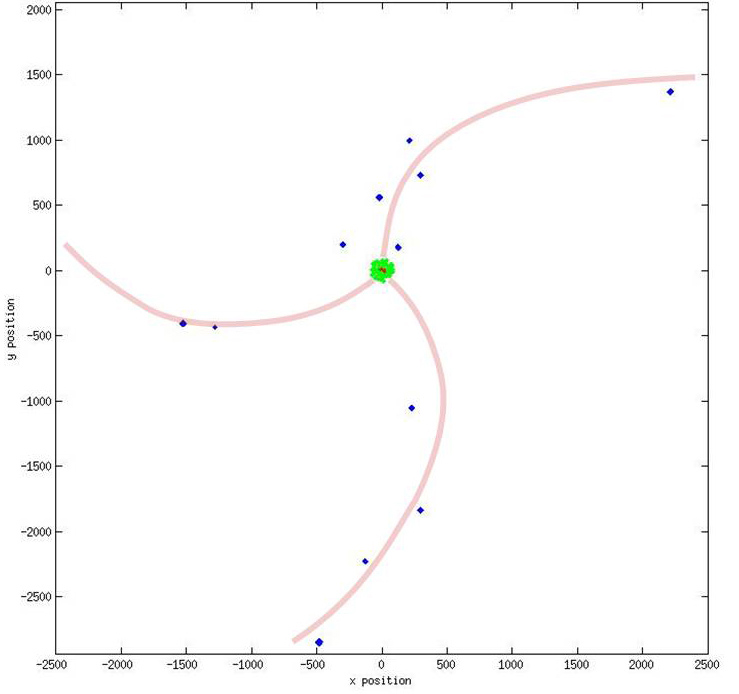}
\caption {Dish configuration showing the middle (blue) and inner
(green) regions with the spiral paths formed for Case 5 when
considering a 24 hour projection.}
\label{dishCase5bPositionsFigure}
\end{figure}

\begin{figure}
\centering
\includegraphics[width=80mm]{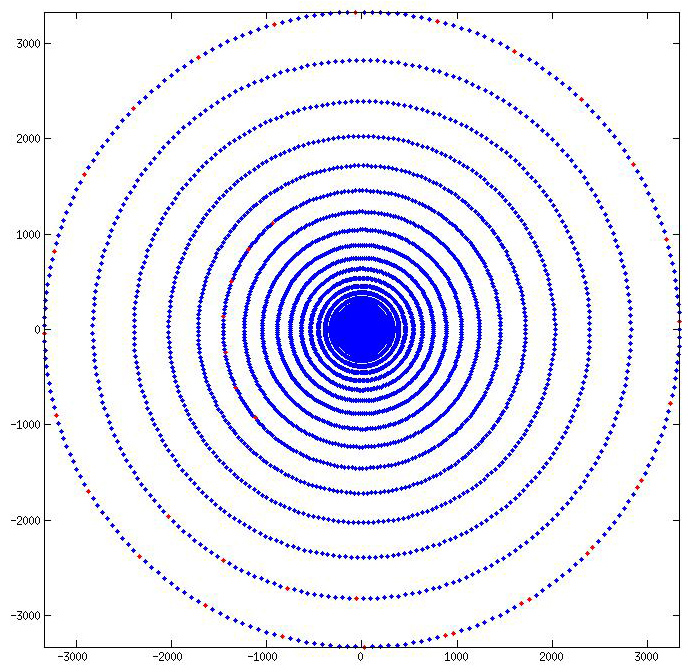}
\caption {Mapping of the 24 hour UV density distribution onto the
nominal grid showing the matched (blue) and unmatched (red) points
for Case 5.} \label{dishCase5bUV}
\end{figure}

To determine how the resolution improves with longer observation
times, another run that takes into account the earth's rotation
and which considers the UV projection over 24 hours, was
performed. Due to the extra calculations involved, the fitness
computation of each chromosome required on average 37.26 seconds.
To finish processing in reasonable time, a population size of 128
was set. The resulting dish positions and the mapping of the UV
points onto the nominal grid are presented in Figure
\ref{dishCase5bPositionsFigure} and Figure \ref{dishCase5bUV}
respectively. As indicated by the shaded tracks, in this case the
GA clearly chose a spiral configuration for the dishes in the
middle region. Moreover, if the three arms are superimposed, the
dishes would be roughly equally spaced along the track. The
attained distribution is causing most of the nominal grid points
to pair after a $360^\circ$ rotation hence giving a very good UV
fitness. The algorithm converged after 101 generations when a good
compromise between UV density and cable length was found.

\subsection{Case 6 - Static SKA CTF and Reuleaux triangle configurations }
\label{dishConfigurationCase6ResultsSubSubSection}

Here, the fitness functions used in the other test cases were
computed for static configurations. In particular, the generic
configuration defined by the SKA Configurations Task Force (CTF)
as well as a dish array specified by Reuleaux triangles were
processed in order to be able to evaluate better the results
achieved by GAs.

\begin{figure}
\centering
\includegraphics[width=80mm]{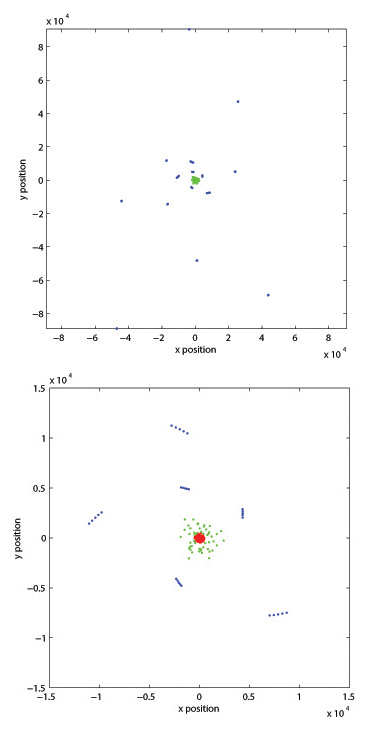}
\caption {Generic SKA CTF dish configuration for Case 6.}
\label{dishCaseGenericPositionsFigure}
\end{figure}

\begin{figure}
\centering
\includegraphics[width=80mm]{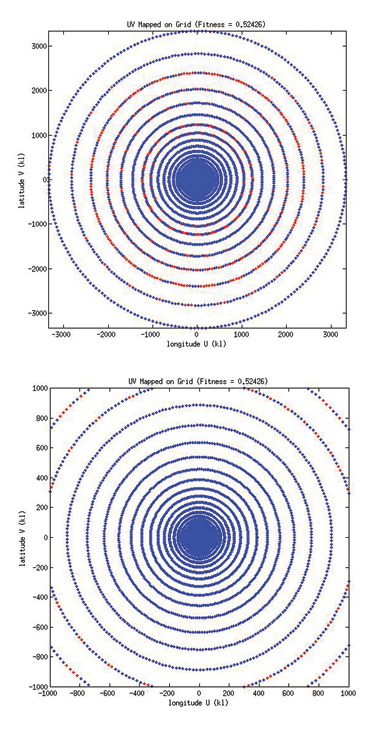}
\caption {Mapping of the 24 hour UV density distribution onto the
nominal grid for the full (top) and core region (bottom) of the
CTF generic array showing the matched (blue) and unmatched (red)
points for Case 6.} \label{dishCaseGenericUvNominal24hourFigure}
\end{figure}

The generic dish configuration by the CTF is shown in Figure
\ref{dishCaseGenericPositionsFigure}. The provided geographical
coordinates were initially converted to cartesian points and
projected onto the regular spatial grid considered in this work.
The UV density distribution was then computed and mapped onto the
nominal grid to obtain the fitness measure $f_{uv}$. This was
found to be 0.8093. The projection of the baseline vector on the
sky over the period of one day was also considered and mapped onto
the nominal grid. As expected this gave better coverage and
reduced $f_{uv}$ to 0.5243. Such a UV mapping is presented in
Figure \ref{dishCaseGenericUvNominal24hourFigure}. The generic
configuration was also processed by the MST algorithm which gave a
wire length of 384.63km. The $f_{WireLog}$, $f_{WireStep}$,
$f_{WirePenalty}$ and $f_{WirePenaltyLow}$ functions resulted to
be 3.5850, 0.05, 0 and 0.5642 respectively.

\begin{figure}
\centering
\includegraphics[width=80mm]{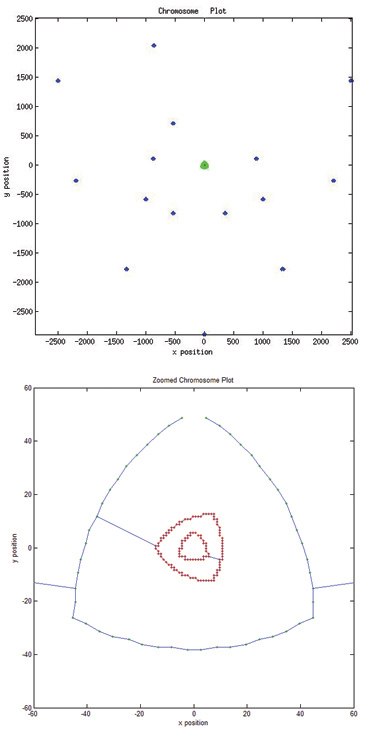}
\caption {Full (top) and zoomed with shortest wire (bottom) dish
configuration showing the middle (blue), inner (green) and core
(red) regions for Case 6.} \label{dishCaseReleauxPositionsFigure}
\end{figure}

After investigating the work by \cite{keto97}, a configuration
defined with Releaux triangles was manually defined and tested.
The dishes in the core were positioned over two slightly rotated
triangles. Similarly, dishes in the inner region were placed
according to a similar but larger shape. Receivers in the middle
region were grouped but still positioned on randomly selected
points from a predefined triangle. The UV fitness from the
baseline vector ($f_{uv}$) was found to be 0.8234. However, when
the 24 hour rotation of the earth was taken into consideration,
this improved to 0.6276. In both cases, the minimum cable length
required to connect all dishes together was found to be
521.8439km. For this case, the $f_{WireLog}$, $f_{WireStep}$,
$f_{WirePenalty}$ and $f_{WirePenaltyLow}$ functions equated to
3.7175, 0.1, 0 and 1 respectively. Figure
\ref{dishCaseReleauxPositionsFigure} shows the defined
configuration and the computed shortest wire for the core and
inner regions.

\subsection{Case 7 - GA with UV, cable length penalty and power spectrum fitness}
\label{dishConfigurationCase7ResultsSubSubSection}

To asses any improvements attained by adding the power spectrum to
the fitness function, a test run with three parameters was set up.
Individuals were ranked by equation \ref{fDish7Eq}.

\begin{equation}\label{fDish7Eq}
f_{dish7} = f_{UV} + f_{WirePenaltyLow} +  f_{PowerSpectrum}
\end{equation}

\begin{figure}
\centering
\includegraphics[width=80mm]{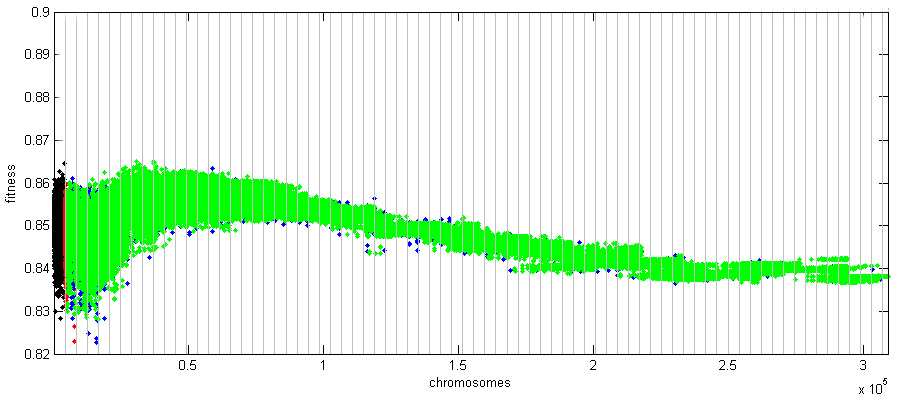}
\caption {Fitness for the initial individuals (black), random
chromosomes (red) and offsprings generated by the mutation (green)
and crossover (blue) operators for Case 7.}
\label{dishCase7FitnessFigure}
\end{figure}

\begin{figure}
\centering
\includegraphics[width=40mm]{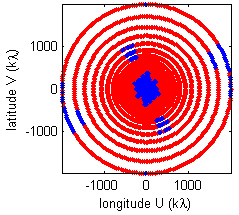}
\caption {UV density distribution showing the matched (blue) and
unmatched (red) points for Case 7.} \label{dishCase7UVFigure}
\end{figure}

\begin{figure}
\centering
\includegraphics[width=80mm]{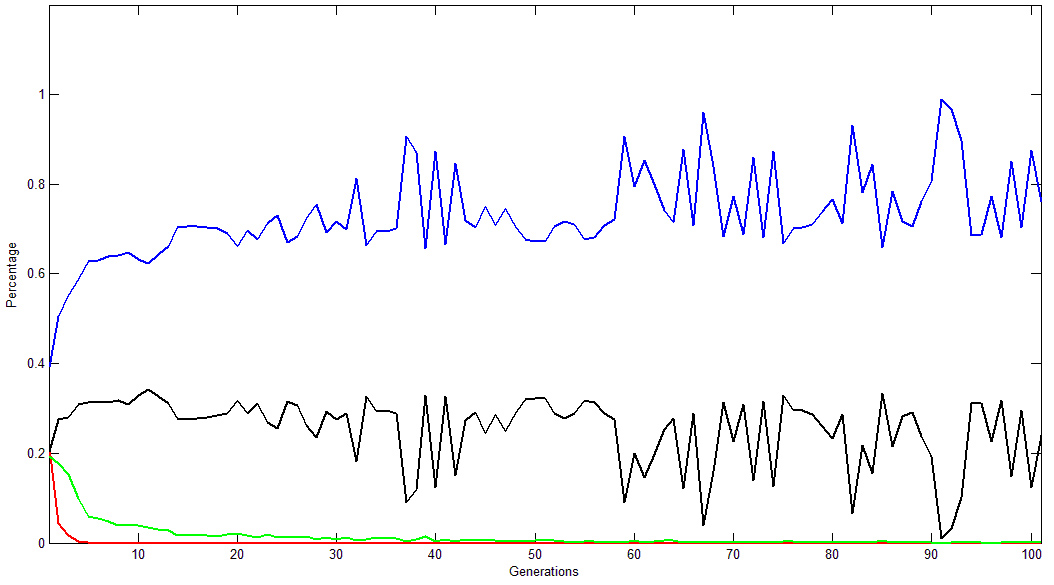}
\caption {Percentage of elite (black), crossover (blue), mutation
(green) and random (red) chromosomes selected for each population
for Case 7.} \label{dishCase7EvolutionFigure}
\end{figure}

In this case, the fitness evaluation function required a few more
milliseconds to processes each chromosome due to the extra
calculations required for spectra computation. The GA still
stalled after about 100 generations. However, a preferred
configuration was determined after the first few iterates and as
shown in Figure \ref{dishCase7FitnessFigure}, no improvement was
attained with subsequent processing. The chosen set of fitness
criteria were not very compatible and hindered the genetic
operators in creating fitter individuals. This can also be seen
from Figure \ref{dishCase7EvolutionFigure} in which constant
migration of chromosomes from one generation to the next is
evident. As can be seen from the UV mapping in Figure
\ref{dishCase7UVFigure}, the dishes in the outer region clustered
together towards the edges and no other configurations were
explored as the algorithm progressed.

The results obtained suggest that no significant improvement is
gained after adding power spectrum estimation to the fitness
function. In particular, the percentage of empty bins in the UV
nominal grid corresponding to the fittest chromosome was found to
be 0.8381\%. The other test runs produced better results.

\section{Results for SKA Phase 2}
\label{dishConfigurationPhase2ResultsSection}

\begin{figure}
\centering
\includegraphics[width=80mm]{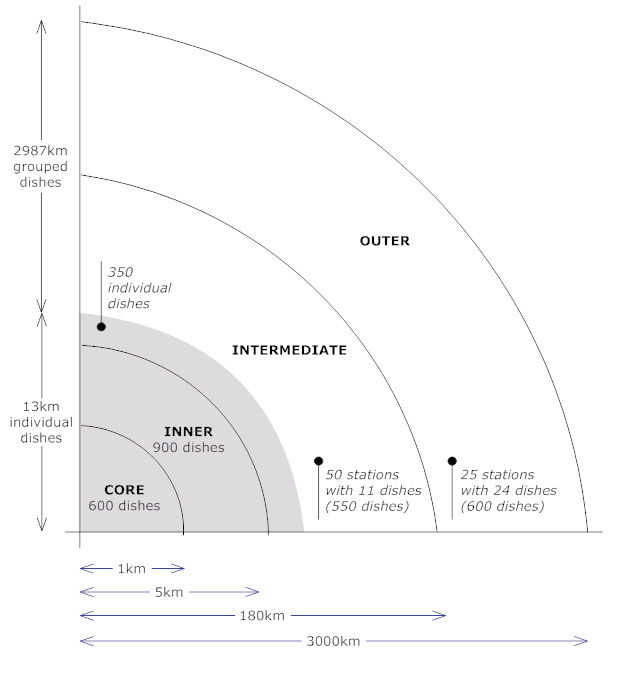}
\caption {SKA Phase 2 dish layout.} \label{dishPhase2LayoutFigure}
\end{figure}

Another genetic algorithm was programmed to search for an optimum
dish configuration solution for SKA phase 2. As defined in
\cite{bolton11}, this will consist of 3000 dishes over a circular
region of up to 3000km in radius. 600 dishes are to be positioned
within the core area of 1km radius and 900 dishes will be
installed up to a radius of 5km from the center. Another 900 and
600 dishes will be set up in the intermediate and outer regions
which are planned to be limited to a 180km and 3000km radius
respectively. Individual dishes will be positioned up to 13km from
the center. Receivers located further away will be grouped. In
this work, the intermediate region was set to contain 350
individual dishes and 50 stations with 11 dishes each. 25 stations
of 24 dishes were set for the outer region. Figure
\ref{dishPhase2LayoutFigure} presents such a dish layout.

Dishes forming part of a group were positioned randomly in a
station whose diameter was considered to vary depending on its
distance from the central core. Specifically, the group radius
$(R_{g})$ was made to vary as defined by equation \ref{rgEq}.

\begin{equation}\label{rgEq}
R_{g} = \sqrt{\frac{log(R) \times A \times N}{\pi}}
\end{equation}

where $R$ is the distance from the central core, $A$ is the area
occupied per dish given by $(pi \times RestZoneRadius)^2$ and $N$
is the number of antennas.

Although the same chromosome structure as that shown in Figure
\ref{dishChromosomeFigure} was used, all genetic operators and
fitness functions had to be redefined due to the different number
of dishes and regions. In order to be able to process and rank the
required number of individuals in adequate time, some assumptions
were made. Dishes in a station where treated as a single point for
UV density calculation. The resulting points were also divided and
mapped onto the nominal grid in parallel. Separate distance
matrices and MSTs were computed for the main regions in order to
obtain an approximate minimum wire length in the shortest time
possible. The $f_{dish3}$ fitness function was used.

\begin{figure}
\centering
\includegraphics[width=80mm]{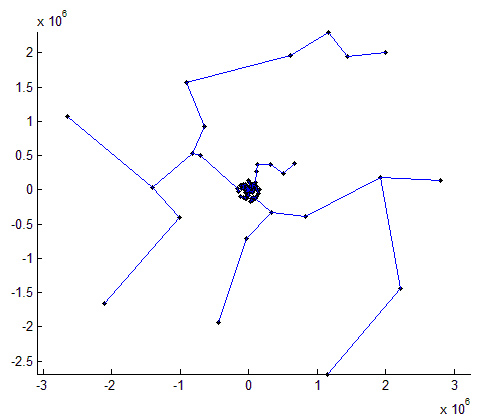}
\caption {Connecting wire for the SKA Phase 2 stations in the
outer region.} \label{dishPhase2WireFigure}
\end{figure}

\begin{figure}
\centering
\includegraphics[width=80mm]{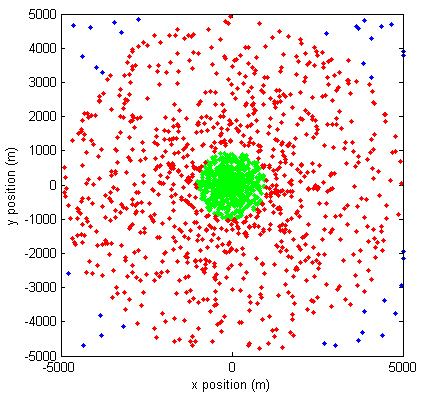}
\caption {Positioning of the SKA Phase 2 dishes in the core
(green) and inner (red) regions. A few receivers in the
intermediate (blue) region can also be seen.}
\label{dishPhase2PositionsFigure}
\end{figure}

\begin{figure}
\centering
\includegraphics[width=80mm]{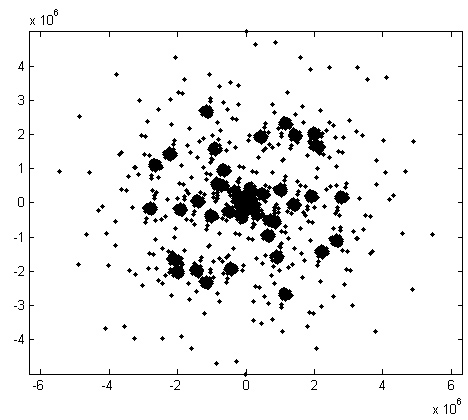}
\caption {UV density distribution for the SKA phase 2 Dishes).}
\label{dishPhase2UvFigure}
\end{figure}

\begin{figure}
\centering
\includegraphics[width=80mm]{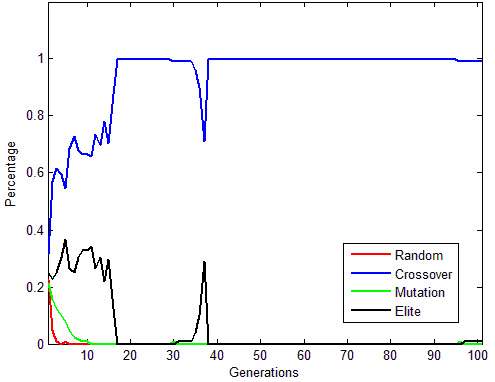}
\caption {Percentage of elite (black), crossover (blue), mutation
(green) and random (red) chromosomes selected for each population
in the SKA Phase 2 run.}\label{dishPhase2EvolutionFigure}
\end{figure}

Figure \ref{dishPhase2WireFigure} shows the wire connecting the
stations in the outer area. Connections between dishes inside a
station are considered negligible. Figure
\ref{dishPhase2PositionsFigure} presents a zoomed plot of the
individual dish locations in the core, inner and intermediate
regions. The resulting UV density plot for the outermost stations
is shown in Figure \ref{dishPhase2UvFigure}. Figure
\ref{dishPhase2EvolutionFigure} gives the percentages of
chromosomes that were generated through crossover, mutation,
randomly or else that migrated from one population to the next
(elite). In this case, the algorithm evolved for 101 generations
before stalling.

\section{Conclusion} \label{conclusionSection}


In this work, we have investigated the use of genetic algorithms
to determine the optimal configuration for the 250 dishes planned
in phase 1 of the SKA telescope as well as for the 3000 dishes
planned for phase 2. The uniformity of uv-distribution and the
connecting wire length were used as parameters for optimisation.
The affects of different dish orientations on the power spectrum,
were also researched.

A number of test cases aimed to investigate different fitness
functions and parameters, were presented. In all experiments,
large genetic population sizes were used as much as possible.
Although an upper limit of 250 was set to the number of
generations, the processes always stalled prior to this and were
stopped when no significant improvement in fitness was detected
with subsequent iterations. In particular, the algorithms always
converged between the 100th and the 120th iterate. The time taken
for each run depended mostly on the fitness functions used. These
were specifically implemented to run in parallel and allowed large
population sizes to be set.

\begin{figure*}
\centering
\includegraphics[width=180mm]{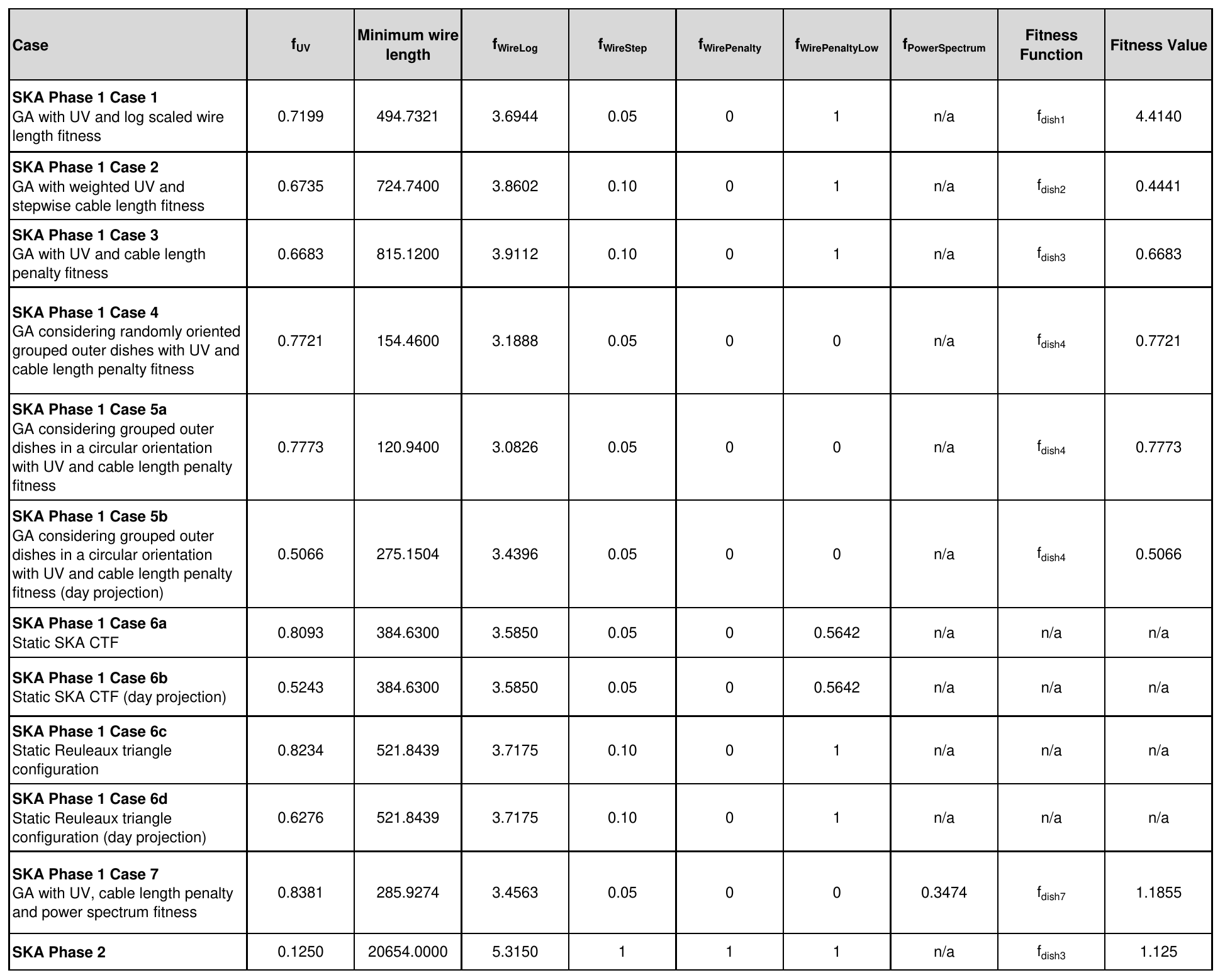}
\caption {Results for the SKA Phase 1 and SKA Phase 2 case
studies.} \label{resultsFigure}
\end{figure*}

A summary of the results obtained in the test cases considered is
presented in Figure \ref{resultsFigure}. Although, the grouping of
the stations in the middle layer improved the wire length fitness,
this had a negative affect on the UV distribution criterion. The
$f_{WirePenaltyLow}$ function was introduced to rank the
individuals correctly even in such cases. This favored chromosomes
that encoded a good tradeoff between UV coverage and cable length.
As expected, better UV sampling was obtained when 24 hour
projections were considered.

Through this and similar work, the potential of machine learning
techniques to aid in identifying optimal dish configurations was
demonstrated. Promising results were obtained and further analysis
can be done once more detailed specifications on the SKA are made
available. For phase 2, the fitness functions had to be slightly
modified and may need to be redefined as the number of dishes and
domain area increase. The work done by \cite{bounova05} which
describes an optimized framework to model robust and scalable
networks, may also be considered to derive the best configuration
for the full SKA.

Future work should also include a more detailed analysis of the
affects of the power spectrum as well as other fitness measures.

\section{Acknowledgements}
\label{acknowledgementsSection} The authors would like to express
their deepest gratitude to Dr Eric R. Keto for his valuable
comments and feedback. His expertise, guidance and support not
only helped to improve this research but have inspired new ideas
for other future work.

The analysis carried out would have not been possible if not for
the computing power made available by the Department of Physics
within the University of Oxford (UK) and the Physics Department
within the University of Malta (Malta).

\bibliographystyle{plainnat}
\bibliography{bibfile}

\label{lastpage}

\end{document}